\documentclass[10pt,showpacs,showkeys,preprintnumbers]{article}
\usepackage{amsmath,amssymb,amsfonts}
\usepackage{units}
\usepackage{bbold,euler}
\usepackage{graphicx}
\usepackage{dcolumn}
\usepackage{bm,bbm,mathtools,esvect}
\usepackage{slashed}
\usepackage{esvect}
\usepackage{fullpage}
\usepackage[utf8]{inputenc}
\title{\LARGE\bf Quaternionic Klein-Gordon equation \vspace{2mm}}
\author{{\bf SERGIO GIARDINO}
\\
\\ Departamento de Matem\'atica Pura e Aplicada,\\ Universidade Federal do Rio Grande do Sul (UFRGS)\\
\\{\tt sergio.giardino@ufrgs.br}
 \vspace{1mm} }

\begin{document}

\maketitle

\begin{abstract}
\noindent We  solve Klein-Gordon equation (KGE) in the framework of the real Hilbert space approach to quaternionic quantum 
mechanics ($\mathbbm H$QM). The presented solution is the simplest ever obtained for quaternionic quantum theories, and the closest
to the complex solution. The scattering of a  quaternionic charged scalar particle from an electric field is also obtained. A remarkable
feature of quaternionic scalar particles is the existence of massive light cone particles.
\end{abstract}



\section{\;\bf Introduction\label{I}}

The  Klein-Gordon equation (KGE) is a quintessential element of quantum field theory, and almost a century of research has not exhausted the activity in the subject. In contrast to the ordered scene observed within the research of quantum field theories based over real and complex functions, the landscape is sharply chaotic among the proposals that generalize quantum fied theory using hyper-complex numbers. Within the ambit of quaternionic quantum field theories ($\mathbbm H$QFT), the broadcasted anti-hermitian approach  recourses to a relativistic scalar product ({\em cf.} Section 11.1 of \cite{Adler:1995qqm}), a resort absent in the well-established complex formulation ($\mathbbm C$QFT). Conversely, there are several attempts to ascertain a mathematically simple and physically understandable approach to a quaternionic Klein-Gordon equation ($\mathbbm H$KGE). The proposals to a $\mathbbm H$KGE are mostly formal, and explicit solutions are scarce. The oldest example uses quaternions to build the space-time structure \cite{Edmonds:1974yq}, but the wave function is not quaternionic. A quaternionic wave function appears initially in modified scalar theories  \cite{DeLeo:1991mi,DeLeo:1995xt,Chanyal:2017xqf}, but also in a fermionic field theory \cite{Ulrych:2013hsq}. In terms of other hyper-complexes, we find a few applications concerning the  KGE made with octonions \cite{Chanyal:2017ull},  sedeonic fields \cite{Mironov:2018bgx,Mironov:2020lnp}, and hyperbolic hyper-complexes \cite{Ulrych:2010ac,Ulrych:2014eoa}. We can educe that the actual investigation status of a generalization for the KGE in terms of hyper-complex numbers comprises isolated proposals, without a physically and mathematically sound framework to support them.

The current scenery is comparable to what was found in non-relativistic quaternionic quantum mechanics ($\mathbbm H$QM) until the year of 2017, when the anti-hermitian formulation of $\mathbbm H$QM  \cite{Adler:1995qqm} was the dominant trend among several quaternionic quantum applications. The main weakness within anti-hermitian $\mathbbm H$QM is the undefined classical limit and the breakdown of the Ehrenfest theorem ({\em cf.} Section 4.4 of \cite{Adler:1995qqm}). In a series of papers, the real Hilbert space approach to $\mathbbm H$QM was introduced, and several consistency tests were achieved, including the Ehrenfest theorem \cite{Giardino:2018lem}, the Virial theorem \cite{Giardino:2019xwm},  and the spectral theorem  \cite{Giardino:2018rhs}. Using this approach, the anti-hermitian requirement of the Hamiltonian operator was removed, and a simpler theory emerged. The  consistency and simplicity of the real Hilbert space approach enabled us to elucidate several unsolved problems of $\mathbbm H$QM, specifically the Aharonov-Bohm effect \cite{Giardino:2016xap},
the free particle \cite{Giardino:2017yke,Giardino:2017nqs}, the square well \cite{Giardino:2020cee}, the Lorentz force \cite{Giardino:2019xwm,Giardino:2020uab}, the quantum scattering \cite{Hasan:2020ekd,Giardino:2020ztf}, and the harmonic oscillator \cite{Giardino:2021ofo}. These results accredit the real Hilbert approach to $\mathbbm H$QM as a promising candidate to generalize quantum mechanics in terms of quaternions.
Therefore, in this paper we intent to apply this successful approach to relativistic $\mathbbm H$QM, an area where quite few investigative work was accomplished, particularly in the case of the Klein-Gordon equation. Therefore, in this paper we prove that the quaternionic KGE can also be solved within the real Hilbert space framework, and we observe that this solution present a close relationship to the complex Klein-Gordon case. However, we observe that the quaternionic solution has their own peculiarities particularly concerning evanescent solutions to KGE, whose existence may be governed or even eliminated from a choice of a proper parameter of quaternionic solutions. The generalized linear momentum also provides quaternionic solutions that deviate from the complex solutions, particurlaly because the Klein paradox can be eliminated without resort to the intensity of the gauge field, but again resorting to the proper quaternionic parameter.
After solving these simplest cases in this paper, we expect that a vast area of applications will appear as future directions of research.

\section{\;\bf Quaternionic Klein-Gordon equation}
Mathematical and physical introductory texts to quaternions ($\mathbbm{H}$) are provided elsewhere \cite{Morais:2014rqc,Rocha:2013qtt,Garling:2011zz,Dixon:1994oqc,Ward:1997qcn}, and in this paper we simply define the necessary notation, recalling that quaternions  are generalized complex numbers that encompass	 three anti-commutative imaginary units, namely $\,i,\,j\,$ and $\,k.\,$ By way of example, the imaginary units satisfy $\,ij=k=-ji.\,$ As a consequence of the anti-commutativity of their imaginary units, quaternionic numbers are non-commutative hyper-complexes. There are several ways of representing quaternions. In the extended notation, that is the na\"ivest notation, all $\,q\in\mathbbm{H}\,$ satisfies
\begin{equation}\label{e01}
 q=x_0 + x_1 i + x_2 j + x_3 k, \qquad\mbox{where}\qquad x_0,\,x_1,\,x_2,\,x_3\in\mathbbm{R},\qquad i^2=j^2=k^2=-1.
\end{equation}
Another useful notation replaces the four real components using a two complex components notation, the symplectic notation. In this setting, (\ref{e01}) becomes
\begin{equation}\label{e02}
q=z_0+z_1j,\qquad\mbox{where}\qquad z_0=x_0+x_1i\qquad\textrm{and}\qquad z_1=x_2+x_3i.
\end{equation}
Several properties of complex numbers have quaternionic analogues, such as the quaternionic conjugate
$\,\overline q,\,$ and the quaternionic norm $\,|q|.\,$ In the symplectic notation, 
\begin{equation}\label{e03}
\overline q=\overline z_0-z_1 j,\qquad\mbox{and}\qquad  |q|^2=q\overline q=|z_0|^2+|z_1|^2.
\end{equation}
Additionally, the  polar quaternionic representation is
\begin{equation}\label{e04}
q=|q|\Big(\cos\theta\, e^{i\varphi}+\sin\theta\, e^{i\xi} j\,\Big).
\end{equation} 
A wave function written in the symplectic notation seems a natural candidate to replace a complex wave function, such that
\begin{equation}
\Psi=\psi^{(0)}+\psi^{(1)} j,
\end{equation}
where $\psi^{(0)}$ and $\psi^{(1)}$ are complex functions, and we expect to recover a complex wave function within the limit $\psi^{(1)}\to 0.$
Let us then entertain the Klein-Gordon equation
\begin{equation}\label{e05}
\Big(\Box+m^2\Big)\Phi=0,\qquad\quad\mbox{where}\qquad\qquad\Box=\partial_\mu\partial^\mu
\end{equation} 
is the usual D'Alembertian operator, that can also be obtained from the quaternionic quantum linear four-momentum for
$\,\hslash=c=1,\,$ so that
\begin{equation}\label{e005}
\widehat{p}_\mu\Phi=\left(\frac{\hslash}{c}\partial_t, -\hslash\bm\nabla\right)\Phi i,\qquad\mbox{so that}\qquad 
-\widehat p_\mu\widehat p^\mu \Phi=\Box\Phi.
\end{equation}
We stress the right hand side position of the imaginary unit $i$ in the operator, as defined in the non-relativistic $\mathbbm H$QM \cite{Giardino:2018lem}.  Additionally, we point out that (\ref{e05}) is not intended to be a sophisticated and mathematically clearer way of expressing the Klein Gordon theory in the complex Hilbert space. Such work has already been done in several physical models \cite{Arbab:2010kr,Arbab:2010dkg}, and we have no contributions to his approach in this article. Conversely, the most important idea underlying (\ref{e05}) is the concept of the real Hilbert space, and the consequent expansion of every quaternionic wave function in a basis of unitary quaternions, as presented in \cite{Giardino:2018rhs}, and not an arrangement of independent real or complex solutions of the KGE within a quaternionic framework. Therefore, following this idea, we propose a quaternionic solution for (\ref{e05}) in terms of the symplectic wave function
\begin{equation}\label{e06}
\Phi=\cos\Theta\,\phi^{(0)}+\sin\Theta \,\phi^{(1)}\,j,
\end{equation}
where $\,\phi^{(0)}\,$ and $\,\phi^{(1)}\,$ are complex functions and $\Theta$ is a real function.  From (\ref{e05}-\ref{e06}), we obtain
\begin{eqnarray}
&&\nonumber\cos\Theta\Big(\Box+m^2-\partial_\mu\Theta\,\partial^\mu\Theta\Big)\phi^{(0)}-\sin\theta\Big(\Box\Theta+2\,\partial_\mu\Theta\,\partial^\mu\Big)\phi^{(0)}+\\
&&+\Big[\sin\Theta\Big(\Box+m^2-\partial_\mu\Theta\,\partial^\mu\Theta\Big)\phi^{(1)}+\cos\theta\Big(\Box\Theta+2\,\partial_\mu\Theta\,\partial^\mu\Big)\phi^{(1)}\Big]j=0,
\end{eqnarray}
that can be rewritten as
\begin{eqnarray}\label{e07}
\Big(\Box	+m^2-\partial_\mu\Theta\,\partial^\mu\Theta\Big)\phi^{(a)}=0\,&&\\
\label{e08}
\,\Big(\Box\Theta+2\,\partial_\mu\Theta\,\partial^\mu\Big)\phi^{(a)}=0\,&&
\end{eqnarray}
where $a=\{0,\,1\}$. Now, we entertain a wave function where
\begin{equation}\label{e09}
\Theta=\theta_\mu x^\mu +\Theta_0,\qquad\mbox{and}\qquad \theta^\mu=\big(\theta_0,\,\bm\theta\big),
\end{equation}
such that $\theta^\mu$ is a constant four-vector and $\Theta_0$ is a constant phase. In this circumstance, $\phi^{(a)}$ will be either real or complex exponential functions, depending whether the constant $\,m^2-\theta_\mu\theta^\mu\,$ is positive, negative or null. Let us choose
\begin{equation}
m^2-\theta_\mu\theta^\mu\geq 0,
\end{equation}
and the quaternionic wave function that solves (\ref{e05}) to be
\begin{equation}\label{e10}
\Phi=\cos\Theta\, \exp\left[ i\,k^{(0)}_\mu x^\mu\right] + \sin\Theta \,\exp\left[i\left( k^{(1)}_\mu x^\mu + \varphi_0\right)\right] j
\end{equation}
where $\varphi_0$ is a constant phase. From (\ref{e07}), a set of constraints also holds, namely
\begin{eqnarray}\label{e11}
 p^{(a)}_\mu p^{(a)\mu}=m^2\qquad \mbox{where}\qquad p^{(a)}_\mu=k^{(a)}_\mu+\theta_\mu,
\end{eqnarray}
and from (\ref{e08}),
\begin{equation}\label{e12}
k^{(a)}_\mu\theta^\mu=0.
\end{equation}
In this situation, we have
\begin{equation}\label{e13}
k_0^{(a)}=\pm\sqrt{m^2+|\bm k^{(a)}|^2-\theta_\mu\theta^\mu\,},
\end{equation}
where $k_0^{(a)}$ is the energy associated to the four-momentum $k_\mu^{(a)}$, although it does not represent the energy of the whole system.  In order to have a better understanding concerning the energy and the momentum of the particle, we propose to entertain the continuity equation for the probability current density four-vector $\mathcal J^\mu$.  We can obtain a null four-divergence  from the real part of the product between equation (\ref{e05})  and $i\overline\Phi$, such that
\begin{equation}\label{e14}
\partial_\mu\mathcal J^\mu=0,\qquad\mbox{where}\qquad 
\mathcal J^\mu=\frac{1}{2m}\Big[\big(\partial^\mu\Phi i\big)\overline\Phi-\Phi i\partial^\mu\overline\Phi\,\Big].
\end{equation}
As an alternative to (\ref{e14}), we have
\begin{equation}\label{e15}
\frac{\partial\,\rho}{\partial t}+\bm{\nabla\cdot\mathit J}=0,
\end{equation}
where the probability density $\,\rho\,$ and the probability current density vector $\,\bm{\mathit J}\,$ are the the components of probability density four-current. Explicitly,  $\,\mathcal J^\mu=\big(\rho,\,\bm{\mathit J}\big),\,$ so that
\begin{equation}\label{e16}
\rho=\frac{1}{2m}\Bigg[\big(\widehat E\Phi\big)\overline\Phi+\Phi \overline{\widehat E\Phi}\,\Bigg],\qquad\mbox{and}\qquad
\bm{\mathit J}=\frac{1}{2m}\Bigg[\big(\widehat{\bm p}\Phi\big)\overline\Phi+\Phi \overline{\widehat{\bm p}\Phi}\,\Bigg],
\end{equation}
and of course
\begin{equation}
\partial_\mu\mathcal J^\mu=0.
\end{equation}
$\,\widehat E\,$ and $\,\widehat{\bm p}\,$ are respectively the energy and momentum operators (\ref{e005}) defined for real Hilbert space $\mathbbm H$QM, so that
\begin{equation}\label{e17}
\widehat E\Phi=\partial_t\Phi i\qquad\mbox{and}\qquad\widehat{\bm p}\Phi=-\bm\nabla\Phi i.
\end{equation}
The quaternionic description of the stationary states of the quaternionic KGE  is formally identical to the complex case, where the probability density $\,\rho\,$ and the probability current density $\,\bm J\,$ have identical interpretations, and the conservation of the probability ascertained in the continuity equation (\ref{e14}). However, in the quaternionic case we have additional degrees of freedom that generate further contributions to these quantities.

Thus, using (\ref{e06}) and (\ref{e14}), we have the general result
\begin{equation}\label{e18}
\mathcal J_\mu=\frac{i}{2m}\Big[\cos^2\Theta\Big(\overline\phi^{(0)}\partial_\mu\phi^{(0)}-\phi^{(0)}\partial_\mu\overline\phi^{(0)}\Big)
-\sin^2\Theta\Big(\overline\phi^{(1)}\partial_\mu\phi^{(1)}-\phi^{(1)}\partial_\mu\overline\phi^{(1)}\Big)\Big].
\end{equation}
The function $\,\Theta\,$ does not contribute directly to the probability current density, and this can be  seen  from the fact that constant $\,\phi^{(a)}\,$ functions lead to an identically zero probability current density, even if $\Theta$ is a non-constant function. Furthermore, using the specific wave function (\ref{e10}), we have
\begin{equation}
\mathcal J_\mu=\frac{1}{m}\Big(-\cos^2\Theta\,k^{(0)}_\mu+ \sin^2\Theta\,k^{(1)}_\mu\Big).
\end{equation}
The above result recover the usual solution of the complex KGE if $\Theta=0$, a wishful characteristic. The $\mu=0$ component shows that negative energies allow negative probability densities, in consonance to the complex cases, where such a problem has been first observed. On the other hand, the energy flux and the the momentum flux oscillate  if $\theta_\mu\neq 0$, and this indicate a more sophisticated system when compared to the complex scalar solutions of the KGE, and the higher number of degrees of freedom in the quaternionic scalar field indicates the wishful expectation that quaternionic quantum theories may describe more complicated systems. Let us see an example from  
\begin{equation}
\mathcal J_\mu\mathcal J^\mu=\frac{1}{m^2}\left(\cos^4\Theta\,k^{(0)}_\mu k^{(0)\mu}+\sin^4\Theta\, k^{(1)}_\mu k^{(1)\mu} -\frac{1}{2}\sin^2  2\Theta\, k^{(0)}_\mu k^{(1)\mu}\right).
\end{equation}
In the complex theory, the above quantity is identically zero for light cone particles. In the quaternionic solution  the expected conditions  $\,k^{(a)}_\mu k^{(a)\mu}=0\,$ are not enough to have a light cone particle. We need the additional constraint
\begin{equation}\label{e20}
k^{(0)}_\mu k^{(1)\mu}=0.
\end{equation} 
Finally, quaternionic light cone particles are admitted to have a non zero mass, in case that $\,\theta_\mu\theta^\mu=m^2.\,$ Massive light cone scalar particles are not allowed in complex KGE, and their existence in the quaternionic theory is of course a remarkable feature.

\section{\bf Generalized quaternionic  Klein-Gordon equation\label{C}      }

In real Hilbert space $\mathbbm H$QM, the generalized linear momentum operator \cite{Giardino:2019xwm} is such that
\begin{equation}\label{g01}
\widehat\Pi_\ell\Phi=-\big(\partial_\ell -\mathcal A_\ell\big)\Phi i,
\end{equation}
where $\,\mathcal A_\ell\,$ is a pure imaginary quaternionic gauge potential, and $\,\ell=\{1,\,2,\,3\}.\,$  In the same fashion as the linear four-momentum (\ref{e005}), the generalized four-momentum operator acts over a quaternionic wave function $\,\Phi\,$ such as
\begin{equation}\label{g02}
\widehat\Pi^\mu\Phi=\big(\partial^\mu -\mathcal A^\mu\big)\Phi i.
\end{equation}
Consequently, the generalized quaternionic Klein-Gordon equation reads
\begin{equation}\label{g03}
\Big(\!\!-\widehat\Pi_\mu\widehat\Pi^\mu+m^2\Big)\Phi=0.
\end{equation}
This equation has the important feature to satisfy a continuity equation analogous to (\ref{e14}), the single difference being only $\partial^\mu\to\Pi^\mu,\,$ such that
\begin{equation}\label{g004}
\partial_\mu\mathscr J^\mu=0,\qquad\mbox{where}\qquad 
\mathscr J^\mu=\frac{1}{2m}\Big[\big(\Pi^\mu\Phi i\big)\overline\Phi+\Phi \overline{\Pi^\mu\Phi i}\,\Big].
\end{equation}
And thus the probability density is conserved in both of the cases, something already observed in non-relativistic $\mathbbm H$QM \cite{Giardino:2016xap,Giardino:2018lem,Giardino:2019xwm}. Let us now entertain the solutions of equation (\ref{g03}), which can be written as
\begin{equation}\label{g04}
\Big(\Box+m^2-|\mathcal A_\mu|^2-\partial_\mu\mathcal A^\mu-2 \mathcal A^\mu\partial_\mu\Big)\Phi=0.
\end{equation}
Considering a solution such as (\ref{e06}) and the quaternionic gauge potential four-vector
\begin{equation}\label{g05}
\mathcal A^\mu=a^\mu i+ b^\mu j
\end{equation}
where $a^\mu$ is a real four-vector and $b^\mu$ is a complex four-vector, we obtain
\begin{eqnarray}
\nonumber &&\cos\Theta\left[\Big(\Box+m^2-\partial_\mu\Theta\,\partial^\mu\Theta-|\mathcal A_\mu|^2-i\partial_\mu a^\mu-2ia_\mu\partial^\mu\Big)\phi^{(0)}+2b_\mu\partial^\mu\Theta\,\overline\phi^{(1)}\right]-\\
\nonumber &&-\sin\Theta\Big[\Big(\Box\Theta+2\,\partial_\mu\Theta\partial^\mu-2ia_\mu\partial^\mu\Theta\Big)\phi^{(0)}-\Big(\partial_\mu b^\mu+2b_\mu\partial^\mu\Big)\overline\phi^{(1)}\Big]+\\
\nonumber &&+\Big\{\sin\Theta\left[\Big(\Box+m^2-\partial_\mu\Theta\,\partial^\mu\Theta-|\mathcal A_\mu|^2-i\partial_\mu a^\mu-2ia_\mu\partial^\mu\Big)\phi^{(1)}+2b_\mu\partial^\mu\Theta\,\overline\phi^{(0)}\right]+\\
\label{g06}
&&+cos\Theta\Big[\Big(\Box\Theta+2\,\partial_\mu\Theta\partial^\mu-2ia_\mu\partial^\mu\Theta\Big)\phi^{(1)}-\Big(\partial_\mu b^\mu+2b_\mu\partial^\mu\Big)\overline\phi^{(0)}\Big]\Big\}\,j=0.
\end{eqnarray}
In the same fashion as (\ref{e07}-\ref{e08}), we obtain
\begin{eqnarray}\label{g07}
&&\Big(\Box+m^2-\partial_\mu\Theta\,\partial^\mu\Theta-|\mathcal A_\mu|^2-i\partial_\mu a^\mu-2ia_\mu\partial^\mu\Big)\phi^{(a)}+2b_\mu\partial^\mu\Theta\,\overline\phi^{(b)}=0\\
\label{g08}
&&\Big(\Box\Theta+2\,\partial_\mu\Theta\partial^\mu-2ia_\mu\partial^\mu\Theta\Big)\phi^{(a)}-\Big(\partial_\mu b^\mu+2b_\mu\partial^\mu\Big)\overline\phi^{(b)}=0.
\end{eqnarray}
Using  (\ref{g004}) and (\ref{e18}), we have the probability density four-current
\begin{eqnarray}\label{g800}
\nonumber\mathscr J^\mu&=&\mathcal J^\mu -\frac{1}{2m}\Big(\mathcal A^\mu\Phi i\overline\Phi+\Phi i\overline\Phi\mathcal A^\mu\Big)\\
&=&\mathcal J^\mu+\frac{1}{m}\Bigg[a^\mu\left(\cos^2\Theta\big|\phi^{(0)}\big|^2-\sin^2\Theta\big|\phi^{(1)}\big|^2\right)
+\frac{i}{2}\sin 2\Theta\Big(b^\mu\overline\phi^{(0)}\overline\phi^{(1)}-\overline b^\mu\phi^{(0)}\phi^{(1)}\Big)\Bigg],
\end{eqnarray}
and both of the components of the gauge potential give contributions to the four-current.
There are many possible ways to solve (\ref{g07}-\ref{g08}), depending on the gauge potential $\mathcal A^\mu$. Let us consider three simple solutions, which are similar to the complex case.
\subsection{\sc  Charged scalar particle in an electric field}
In this case, let us choose a pure complex quaternionic electric potential, such that
\begin{equation}
a_\mu=a_0,\qquad a_\ell=0,\qquad b_\mu=0,
\end{equation}
and $\,a_0\,=eV_0,$ where $e$ is the electric charge and $V_0$ is a constant electric potential. In this case, adopting the positive signal in the argument of the exponential functions of (\ref{e10}), equations
(\ref{g07}-\ref{g08}) give
\begin{eqnarray}
\label{g09} k^{(a)}_\mu k^{(a)\mu}&=&m^2-a_0^2+2a_0k_0-\theta_\mu\theta^\mu\\
\label{g10} a_0\theta_0&=&\theta_\mu k^{(a)\mu}.
\end{eqnarray} 
Using $\,p^{(a)\mu}=k^{(a)\mu}+\theta^\mu\,$ from (\ref{e11}), we get
\begin{equation}
	p_\mu^{(a)}p^{(a)\mu}=2a_0(\theta_0+k_0)-a_0^2+m^2
\end{equation}
In terms of the linear momentum, we have
\begin{equation}\label{g100}
\big|\bm p^{(a)}\big|^2=\left(p^{(a)}_0-a_0\right)^2-m^2.
\end{equation}
The analogy to the complex case is simply perfect, we have two effective momenta and both of them satisfy relations that are equivalent to that found in the complex KGE.
\subsection{\sc  Charged scalar particle in an constant quaternionic gauge field}
Let us choose a pure quaternionic four-potential, such that $a^\mu=0$.
 Adopting the (\ref{e10}) wave function, (\ref{g07}-\ref{g08}) give
\begin{eqnarray}\label{g10}
&&\Big(-k_\mu^{(a)}k^{(a)\mu}+m^2-\theta_\mu\theta^\mu-b_\mu\overline b^\mu\Big)\phi^{(a)}+2b_\mu\theta^\mu\,\overline\phi^{(b)}=0,\\
\label{g11}
&&\theta_\mu\, k^{(a)\mu}\,\phi^{(a)}+b_\mu\, k^{(b)\mu}\,\overline\phi^{(b)}=0,
\end{eqnarray}
where $\,b^\mu\,=b^{(0)\mu}+b^{(1)\mu}i$. In the simplest case,
\begin{equation}\label{g11}
\theta^\mu=0\qquad\Rightarrow\qquad\left\{
\begin{array}{l}
q_\mu^{(a)}\overline q^{(a)\mu}=m^2\\
b_\mu\, k^{(a)\mu}=0
\end{array}
\right.,\qquad
\mbox{where}\qquad q^{(a)\mu}=k^{(a)\mu}+b^\mu.
\end{equation}
As a result, the net effect of the $\,b^\mu\,$ complex gauge field is to define  the effective complex momenta $\,q^{(a)\mu},\,$ that fit the mass and the charge of the particle, and the final result is thus similar to the complex case, and the resulting quaternionic particle is composed of two coupled complex particles. A generalization of the above result is such that
\begin{equation}\label{g12}
\phi^{(a)}={\overline\phi}^{(b)},\qquad\mbox{or equivalently,}\qquad -k^{(b)\mu}=k^{(a)\mu}=k^\mu
\end{equation}
and consequently
\begin{equation}
\left\{
\begin{array}{l}
\theta_\mu b^\mu=0\\
\big(\theta_\mu+b_\mu\big)k^\mu=0
\end{array}
\right.
\qquad\Rightarrow\qquad q_\mu\overline q^\mu=m^2-\theta_\mu\theta^\mu,
\end{equation}
and $\theta^\mu=0$ recovers (\ref{g11}). Finally, let us consider another possibility for (\ref{g12}), where 
\begin{equation}
b^{(0)}_0=b_0,\qquad b^{(0)}_\ell=0,\qquad b_\mu^{(1)}=0,
\end{equation}
and $\,b_0=eV_0\,$ is constant. Thus, we obtain
\begin{eqnarray}\label{g13}
k_\mu k^\mu&=&m^2-\theta_\mu\theta^\mu-b_0^2+2bo\theta_0\\
b_0 k_0&=&\theta_\mu k^\mu.
\end{eqnarray}
Using $\,p^\mu=k^\mu+\theta^\mu\,$ Manipulating (\ref{g13}) we obtain
\begin{equation}
|\bm p|^2=\big(p_0-b_0\big)^2-m^2.
\end{equation}
and the correspondence to (\ref{g100}) is exact, and the single difference is that $\theta^\mu=0$ leads to a trivial solution, something that does not happen in the previous case.  
\subsection{\sc  Charged scalar particle in an oscillating quaternionic gauge field}
Let us suppose that
\begin{equation}
\phi^{(0)}=\phi^{(1)},\qquad\mbox{so that}\qquad k^{(a)\mu}=k^\mu.
\end{equation}
In this case, let us solve (\ref{g08}) using
\begin{equation}
a^\mu=0,\qquad b^\mu=\beta^\mu \exp\big[2ik_\mu x^\mu\big] \qquad\mbox{and}\qquad \theta_\mu k^\mu=0,
\end{equation}
where $\beta^\mu$ is a constant real four-vector. Consequently, 
\begin{equation}
k_\mu k^\mu=m^2-\big(\theta_\mu-\beta_\mu\big)\big(\theta^\mu-\beta^\mu\big).
\end{equation}
This solution, however, has a peculiar feature. From (\ref{g800}), the probability density current reads
\begin{equation}
\mathscr J^\mu=-\frac{1}{m}\cos 2\Theta\, k^\mu.
\end{equation}
Imposing the additional constraints 
\begin{equation}
k_\mu k^\mu=0\qquad\mbox{and}\qquad m^2=\big(\theta_\mu-\beta_\mu\big)\big(\theta^\mu-\beta^\mu\big),
\end{equation}
we have a massive light cone particle, a quaternionic attribute already observed in (\ref{e20}), but unknown in
the complex case. The results of this section confirm that the $\mathbbm H$KGE contain solutions that are fundamentally different from that found in the complex case, and can therefore be considered a quantum theory of higher generality.
\section{\sc Scattering of a charged quaternionic scalar particle \label{C}      }

In this final section, we study the scattering of a quaternionic charged scalar particle from the constant one-dimensional step gauge potential four vector
\begin{equation}\label{s01}
\mathcal A^\mu=
\left\{
\begin{array}{cl}
0& \qquad\mbox{if}\qquad  x\leq 0\\
\mathcal A^\mu&\qquad\mbox{if}\qquad  x>0,
\end{array}
\right.
\end{equation}
where $\,\mathcal A^\mu\,$ is a pure imaginary constant quaternion.
Let us consider the solution
\begin{equation}\label{s02}
\Phi=
\left\{
\begin{array}{ll}
\Phi_I=\Phi^{(0)}+ R\, \Phi^{(1)}& \qquad\mbox{if}\qquad  x\leq 0\\
\Phi_{II}=T\,\Phi^{(2)} &\qquad\mbox{if}\qquad  x>0.
\end{array}
\right.
\end{equation}
where $\,R\,$ and $\,T\,$ are complex constants, and the components of the wave function are
\begin{equation}\label{s03}
\Phi^{(s)}=\cos\Theta\exp\big[ip^{(s)}_\mu x^\mu\big]+\sin\Theta\exp\big[iq^{(s)}_\mu x^\mu\big] \,j,
\end{equation}
where $s=\,\{0,\,1,\,2\}\,$ and (\ref{e09}) holds. The wave function and their first derivative are not both continuous at the $\,x=0\,$ scattering point, because there are too many constraints to be satisfied, and consequently the $\mathbbm C$KG solution is recovered, and nothing new appears. Such kind of difficulty has already been observed in  the case of non-relativistic scattering of by rectangular potentials \cite{Giardino:2020cee}. Let us then impose the continuity of the probability four-current, such that
\begin{eqnarray}\label{s04}
\nonumber \mathcal J_I^\mu=\frac{1}{m}\Bigg\{\!\!&-&\cos^2\Theta\Big(p^{(0)\mu}+|R|^2\, p^{(1)\mu}\Big)+\sin^2\Theta\Big(q^{(0)\mu}+|R|^2 \,q^{(1)\mu}\Big)+\\
&+&|R|
\Bigg[-
\Big(p^{(0)\mu}+p^{(1)\mu}\Big)\cos^2\Theta\,\cos\Big[\big(p^{(0)}_\mu-p^{(1)}_\mu\big)x^\mu-\phi_R\Big]+\\
\nonumber&+&
\Big(q^{(0)\mu}+q^{(1)\mu}\Big)\sin^2\Theta\,\cos\Big[\big(q^{(0)}_\mu-q^{(1)}_\mu\big)x^\mu-	\phi_R\Big]
\Bigg]\Bigg\}.
\end{eqnarray}
where we used $\,R=|R|e^{i\phi_R}.\,$ Furthermore,
\begin{equation}\label{s05}
\mathcal J^\mu_{II}=\frac{|T|^2}{m}\Big(-p^{(2)\mu}\cos^2\Theta+q^{(2)\mu}\sin^2\Theta \Big)-\frac{1}{2m}\Big(\mathcal A^\mu\Phi^{(2)} i\overline\Phi^{(2)}+\Phi^{(2)} i\overline\Phi^{(2)}\mathcal A^\mu\Big).
\end{equation}
Let us choose
\begin{eqnarray}\label{s06}
&&p^{(0)\mu}=\big(-p_0,\,-p_\ell\big),\qquad p^{(1)\mu}=\big(-p_0,\,p_\ell\big),\qquad p^{(2)\mu}=\big(-p_0,\,-P_\ell\big),\\
&&q^{(0)\mu}=\big(-q_0,\,q_\ell\big),\qquad q^{(1)\mu}=\big(-q_0,\,-q_\ell\big),\qquad q^{(2)\mu}=\big(-q_0,\,Q_\ell\big).\\
\end{eqnarray}
The continuity of the probability current at $x_\ell=0$ give
\begin{equation}\label{s07}
1+|R|^2+2\cos\phi_R=|T|^2,\qquad 1-|R|^2=\beta |T|^2\qquad\mbox{where}\qquad \beta=\frac{P_\ell}{p_\ell}=\frac{Q_\ell}{q_\ell}.
\end{equation}
The first equation allow us to state that
\begin{equation}\label{s08}
|T|^2=\big(1+R\big)\big(1+\overline R\big)\qquad\Rightarrow\qquad 1+R=T e^{i\delta},
\end{equation}
where $\,\delta\,$ is an arbitrary phase. Using (\ref{s08}) to calculate $|R|^2$, we obtain that
\begin{equation}\label{s09}
T=|T|e^{i\phi_T},\qquad |T|=\frac{2}{1+\beta}\cos\big(\phi_T+\delta\big),\qquad\mbox{and}\qquad R=\frac{e^{2i(\phi_T+\delta)}-\beta}{1+\beta}
\end{equation}
In the $\phi_T+\delta=0$ case, we obtain the usual coefficients to complex scattering (Section 1.3.1 of \cite{Das:2008zze}). From the spatial component of the four-current, we obtain reflection and transmission coefficients, we have
\begin{equation}
\mathcal J^\ell_{INC}=\frac{p_\ell}{m}\cos^2\Theta, \qquad \mathcal J^\ell_{REF}=\frac{p_\ell}{m}|R|^2\cos^2\Theta,\qquad 
\mathcal J_{TRANS}^\ell=\frac{P_\ell}{m}|T|^2\cos^2\Theta.
\end{equation}
Finally, defining the transmission and reflection coefficients,
\begin{equation}
\mathcal R =\frac{J^\ell_{REF}}{J^\ell_{INC}}=|R|^2,\qquad\mbox{and}\qquad \mathcal T =\frac{J^\ell_{TRANS}}{J^\ell_{INC}}=\beta |T|^2,
\end{equation}
we immediately obtain from (\ref{s07})
\begin{equation}
\mathcal R+\mathcal T=1.
\end{equation}
We observe that the physical features of the conservation of probability is also satisfied in the $\mathbbm H$KGE. Addtionally, the Klein paradox, where a negative reflection coefficient is obtained for negative $\,\beta,\,$ and thus the particle generation may be obtained inverting the signal of the momentum of the pure quaternionic component of the wave function. 

\section{\sc Conclusion \label{C}      }

In this article we solved the $\mathbbm H$KGE, and this is the simplest solution ever obtained. The wave function comprises  complex and  quaternionic  components that have independent energies and momenta. These combined solutions have new features, like massive and nonphotonic light cone particles, something that is not allowed in the $\mathbbm C$KGE. Thus, the additional degrees of freedom of the quaternionic wave functions enable the existence of physical situations that are   unknown in complex case, and open the possibility to seek and explain new phenomena.  The presented results reinforce the capacity of the real Hilbert space approach of $\mathbbm H$QM to produce novel and consistent results. There is a broad pathway that can be explored as future directions of research. These directions include every application of the $\mathbbm C$KGE, and also the challenging research involving  the construction of a consistent quaternionic scalar field theory,  where the lagrangian and hamiltonian formulations have to be presented in full detail. We can anticipate that this theory is in a final stage of developtment, and will be published soon.

%
%
%
%

\bibliographystyle{unsrt} 
\bibliography{bib_KG}
\end{document}